\documentclass[aps,prl,reprint,superscriptaddress,,nofootinbib,
showpacs]{revtex4-1}
\usepackage{amsmath,amssymb,bm,float,mathrsfs}
\usepackage{graphicx}
\usepackage{dcolumn}
\usepackage{color}
\usepackage{hyperref}
\usepackage{comment}
\usepackage{fancybox}

  \newcommand{\beq}{\begin{equation}}
  \newcommand{\eeq}{\end{equation}}
  
  \newcommand{\bi}{\begin{itemize}}
  \newcommand{\ei}{\end{itemize}}

\begin{document}

\title{Evidence for Gauge/Gravity Correspondence for D$p$-branes\\
at Weak 't Hooft Coupling}

\author{Yasuhiro Sekino}
\affiliation{Department of Liberal Arts and Sciences,
Faculty of Engineering,
Takushoku University, Tokyo 193-0985, Japan}

\begin{abstract}
We consider gauge/gravity correspondence for general D$p$-branes
with $0\le p\le 4$,
namely, the duality between the $(p+1)$-dimensional maximally 
supersymmetric Yang-Mills theory and 
superstring theory on the near-horizon limit of
the D$p$-brane solution. 
It has been known that in the large $N$ gauge theory at zero temperature 
and strong 't Hooft 
coupling, the two-point functions of operators corresponding 
to supergravity modes obey power-law, with a power different 
from the free-field value when $p\neq 3$. 
In this work, we show that the free-field result of gauge theory 
can be reproduced from the bulk string theory, for the two-point function of
an operator ${\rm Tr}Z^J$, where $Z$ is a complex combination
of two scalar fields. We assume the spatial direction of worldsheet 
is discretized into $J$ bits, and use the fact that these bits become
independent particles when string tension is zero.
\end{abstract}






\maketitle

\paragraph*{Introduction.---}
Gauge/gravity correspondence~\cite{Maldacena} 
is a proposal for concrete realizations
of the holographic principle~\cite{Susskind}. 
Quantum gravity in various spacetimes 
are conjectured to be equivalent to gauge theories
defined at the spatial boundary. 
Gauge/gravity correspondence is now being applied to the theories beyond
the original proposals based on string/M-theory, in such areas as 
nuclear physics and condensed matter physics. (See e.g.\ \cite{Ammon,
Liu, Zaanen, Hartnoll} for reviews.)
It is widely regarded
as a powerful theoretical tool which allows one to study strongly
coupled quantum theory by gravity that is simple.

Gauge/gravity correspondence has not been proven yet, and we do not
know how (or why) it works. An unresolved issue that
should be closely related to the proof
is what kind of correspondence there is when gauge theory 
is weakly coupled. There are very few studies\footnote{%
For the studies of the same SYM theories as the 
one studied here, see
\cite{Wadia, Karch, Gopakumar1, Gopakumar2, Gopakumar3, 
BerkovitsVafa, Berkovits1, Berkovits2, Alday, Nastase}.
There have been studies of somewhat different theories
based on higher-spin symmetry: For AdS$_3$/CFT$_2$,
see e.g.\ \cite{Gaberdiel1, Gaberdiel2} and references therein;
for AdS$_4$/CFT$_3$ involving 3D vector models, see 
e.g\ review articles \cite{Yin, Jevicki}.}
on this limit, compared
to the active studies on the other strong coupling limit. 
The purpose of this Letter is to point out an intriguing fact
about this limit. 

We consider gauge/gravity correspondence associated with
the two descriptions of D$p$-branes,
namely the conjectured equivalence between the
$(p+1)$-dimensional  maximally supersymmetric $SU(N)$ Yang-Mills 
theory and superstring theory on the near-horizon limit of the D$p$-brane 
solution~\cite{IMSY}. The former is the low-energy effective theory
on the D$p$-brane worldvolume based on the open-string lowest mode, 
and the latter gravity description is based on the closed-string 
degrees of freedom. 

The $p=3$ case is the well-studied 
AdS$_5$/CFT$_4$ correspondence between the (3+1)dimensional ${\cal N}$=4 
supersymmetric $SU(N)$ Yang-Mills theory and superstring theory 
on AdS$_5\times S^5$~\cite{Maldacena}. 
We will consider the general case with $0\le p\le 4$.
For $p\neq 3$, there is no conformal invariance, and 
exact results are hard to obtain, but 
there have been quantitative studies (especially for the $p=0$ case) 
\cite{Hashimoto, SekinoYoneya, Sekino, ASY, AS, Asano, 
HNSY1, HNSY2, Oz, Wiseman, Skenderis, Ortiz}. Correlation functions at strong 't Hooft coupling
have been obtained by tree-level supergravity. 
The results for $p=0$ have been confirmed by Monte Carlo
simulations~\cite{HNSY1, HNSY2}, providing strong evidence
for the gauge/gravity correspondence without conformal symmetry. 

We will consider gauge theory at zero temperature, and 
study the two-point function of a single-trace operator with 
large angular momentum $J$ on the transverse $S^{8-p}$, which 
has been introduced by Berenstein, Maldacena and Nastase 
(BMN)~\cite{BMN} in the study of strings on the plane wave background. 
We show that the free-field result of gauge theory can be reproduced
by string theory, by making an assumption that string
is composed of $J$ ``bits,'' which is natural in view of
previous work~\cite{BMN, Verlinde1, Verlinde2, Zhou,
Bellucci, Danielsson, Alday}. 
In the following, we will first review gauge/gravity correspondence
for general $p$ at strong 't Hooft coupling, to set
the stage for weak 't Hooft coupling. The main result 
is presented near the end.

\paragraph*{The background.---}
The near-horizon limit of the metric and the dilaton for 
the zero-temperature D$p$-brane solution in the string frame are 
\begin{eqnarray}
ds^{2}&=&H^{-1/2}\left(-dt^{2}+d x_{a}^{2}\right)
+H^{1/2}\left(dr^{2}+r^{2}d\Omega_{8-p}^{2}\right),
\nonumber\\
e^{\phi}&=& g_{s}H^{\frac{3-p}{4}},
\quad H = {q\over r^{7-p}} 
\label{eq:nhDp}
\end{eqnarray}
where $a=1,\ldots, p$ and 
$q=\tilde{c}_{p}g_{s}N\ell_{s}^{7-p}$ with 
$\tilde{c}_{p}=2^{6-p}\pi^{(5-p)/2}\Gamma{(7-p)/2}$. 
The integer $N$ is the number of the D$p$-branes,
$\ell_s$ is the string length.
The Yang-Mills coupling is given 
by $g^2_{\rm YM}=(2\pi)^{p-2}g_s\ell_s^{p-3}$.

For $p=3$, the near-horizon geometry is AdS$_5\times S^5$; 
for general $p$, it is related to 
AdS$_{p+2}\times S^{8-p}$ by a Weyl rescaling,
\begin{equation}
ds^2 =H^{1/2}r^2 
\left[
\left({2\over 5-p}\right)^{2}\left( dt^{2}+dx_{a}^{2}+dz^{2}\over z^{2}\right)
+d\Omega_{8-p}^2\right]
\label{eq:Weyl}
\end{equation}
where the radial variable $z$ in the Poincar\'{e} coordinates
for AdS$_{p+2}$ is defined by
\begin{equation}
z={2\over 5-p}(g_sN)^{1/2}\ell_s^{(7-p)/2}r^{-(5-p)/2}.
\end{equation}
The boundary is at $z\to 0$. The distance $|\Delta x|$
in gauge theory roughly corresponds to the region 
$z\lesssim |\Delta x|$~\cite{SusskindWitten, PeetPolchinski}, as we can
see e.g.\ in the geodesic approximation. 

For $p\neq 3$, the dilaton and the curvature depends on the radial
position, and the background do not have the AdS isometries; 
correspondingly, gauge theory does not have conformal invariance. 
Nevertheless, the representation \eqref{eq:Weyl} is useful, since  
null geodesics are not affected by the Weyl factor, supergravity 
modes, which are massless in 10 dimensions, shows similar behavior
(obeying power-law) to the conformal case at the tree level.

In this work, we always assume the string coupling is 
small, $e^{\phi}\ll 1$, so that we can ignore 
string loop effects. For $p<3$,
this is satisfied in the near-boundary (UV) region,
$z\ll (g_sN)^{-1/(3-p)}N^{2(5-p)/((7-p)(3-p))}\ell_s$,
and for $p>3$, satisfied in the center (IR) region,
$(g_sN)^{-1/(3-p)}N^{2(5-p)/((7-p)(3-p))}\ell_s\ll z$.
If curvature is small in string units, string higher excitations can
be ignored. For $p<3$, this is satisfied in the IR region,
$(g_sN)^{-1/(3-p)}\ell_s\ll z$, and for $p>3$, satisfied in 
the UV region, $z\ll (g_sN)^{-1/(3-p)}\ell_s$. 
If we take $N\to \infty$ with the 't Hooft coupling fixed but 
large $g_s N\gg 1$, the above two conditions are satisfied 
in almost the whole near-horizon region~\cite{SekinoYoneya}. 
In that case, the tree-level supergravity
is expected to be valid. When we take 't Hooft coupling to be
small, we should consider strongly curved background. 

\paragraph*{The operator.---}
We will focus on the two-point functions of a
``BMN operator''~\cite{BMN},
\begin{equation}
{\cal O}={\rm Tr} \left( Z^{J}\right)
\label{eq:OZJ}
\end{equation}
where $Z=X^{8}+iX^{9}$ is a complex combination of two of the $(9-p)$
scalar fields (that arise from dimensional reduction of the 
(9+1)-dimensional gauge field)
in the $(p+1)$-dimensional maximally supersymmetric 
Yang-Mills theory. The integer $J$ corresponds to a quantum number 
for the $SO(9-p)$ symmetry (angular momentum along a great sphere 
in the $X^8$-$X^9$ plane). 

For $p=3$, this operator is one of the BPS operators, which
belong to a short representation of the superconformal algebra.
Its scaling dimension is given by the free-field value 
$\Delta=J$, and is protected against quantum corrections. 
The BPS operators correspond to supergravity modes. 

For $p\neq 3$, the analogs of the BPS operators, which are 
related to the ones for $p=3$ by "T-duality" (or dimensional
reduction), couple to supergravity modes~\cite{TR1,TR2, TR3}.
The full spectrum
of the supergravity modes has been obtained for the D0-brane 
background~\cite{SekinoYoneya, Sekino}, and the corresponding operators 
in the $(0+1)$D SYM have been identified~\cite{SekinoYoneya, Sekino}
with a help of the ``generalized conformal 
symmetry''~\cite{JevickiYoneya}\footnote{%
This is a symmetry realized when the coupling constant is
allowed to vary. To the author's knowledge, this symmetry does
not have a power to algebraically constrain the scaling w.r.t.\ the 
coordinate separation, unlike the true superconformal symmetry.}. 
The operator \eqref{eq:OZJ} belongs to $T^{++}_{J}$ defined in \cite{TR1},
and corresponds to the supergravity mode called 
$s^{3}_{J}$ in \cite{SekinoYoneya}. 

\paragraph*{Supergravity analysis.---}
We assume the general relation due to Gubser, Klebanov, Polyakov~\cite{GKP}
and Witten~\cite{Witten} (GKPW), between the bulk partition 
function $Z[\phi_0]$ and the generating functional for correlation
functions in gauge theory, holds,
\begin{equation}
Z[\phi_0]=\langle e^{\int d^{p+1}x \phi_{0}(x) 
{\cal O}(x)}\rangle.
\label{eq:GKP}
\end{equation} 
Here, $\phi_0$ is the boundary condition for a bulk field $\phi$,
imposed at the $(p+1)$-dimensional boundary of the AdS$_{p+2}$-like space, 
and ${\cal O}$ is the operator which couples to 
$\phi_0$. Calculations are
performed in the Euclidean signature. 
In the limit of weak string coupling, the bulk partition function 
is given by the classical action,
$
Z[\phi_0]= e^{-S_{\rm SG}[\phi_0]},
$
and can be calculated by the tree-level supergravity using 
bulk-to-boundary propagators. 
By applying the GKPW prescription to the near-horizon D$p$-brane 
background, the two-point function of the operator \eqref{eq:OZJ} 
has been found to be~\cite{SekinoYoneya}  
\begin{equation}
\langle{\cal O}(x'){\cal O}(x)\rangle
={\delta \over \delta\phi_0(x')}{\delta \over \delta\phi_0(x)}
e^{-S_{\rm SG}[\phi_0]}\sim {1\over |x-x'|^{{4J\over 5-p}+c_p}}.
\label{eq:OOGKPW}
\end{equation}
where we have ignored an overall constant factor. The constant $c_p$ in the 
exponent takes the value\footnote{%
This value of $c_p$ has been deduced from the available supergravity 
results for $p=0$ and $p=3$.}
\begin{equation}
c_p=-{(3-p)^2\over 5-p}.
\label{eq:c}
\end{equation}

The correlator \eqref{eq:OOGKPW} obeys power law\footnote{%
Correlation functions for operators corresponding to 
string higher modes are predicted to be exponential 
functions~\cite{ASY, Asano}.}, 
even though the coupling constant 
$g_{\rm YM}$ has a dimension for $p\neq 3$.
The power is different from the free-field value 
for $p\neq 3$. 
Strong coupling dynamics together with supersymmetric
cancellations should be responsible 
for this behavior in gauge theory, 
but the mechanism is not understood analytically yet. 
For $p=0$, this power law
has been confirmed by Monte Carlo 
simulation~\cite{HNSY1, HNSY2}.

\paragraph*{Geodesic approximation.---}
Supergravity modes are massless in 10D spacetime, but when 
we regard them as Kaluza-Klein modes, angular momentum $J$ 
on $S^{8-p}$ corresponds to mass\footnote{%
For Kaluza-Klein reduction of massless particle on \eqref{eq:Weyl},
the Weyl factor plays no role, and  
we effectively obtain a massive particle action 
on the true AdS$_{p+2}$~\cite{ASY, AS}.
The factor $2/(5-p)$ is due to the ratio 
of the radii of AdS$_{p+2}$ and $S^{8-p}$~\cite{ASY,AS}.} 
$m={2\over 5-p}J$ in AdS$_{p+2}$. 
In the large $J$ limit, the geodesic approximation can be used
to obtain the correlator. 
 
We will consider Euclidean AdS$_{p+2}$. 
Then, a geodesic of a massive particle connects $x_i$ and
$x_f$ on the boundary (separated e.g.\ in the Euclidean time direction). 
We will regulate the infinite volume of the AdS$_{p+2}$, and put
the boundary at the radial position $z=1/\Lambda$. 
The geodesic is a half sphere $x^2+z^2=\tilde\ell^2$
in the Poincar\'{e} coordinates. In terms of the proper time $\tau$,
\begin{equation}
z={\tilde\ell\over \cosh\tau}, \quad x=\tilde\ell \tanh\tau,
\label{eq:zt}
\end{equation}
where $\tilde\ell$ is a parameter corresponding to the distance 
of the two points along the boundary $|x_i-x_f|=2\tilde\ell$. 
The geodesic reaches the boundary at an infinite proper time, thus
we introduce a cutoff $-T\le \tau\le T$. The relation between
the cutoffs $\Lambda$ and $T$ is given from \eqref{eq:zt} as
$e^{T}\sim 2\tilde\ell\Lambda=|x_f-x_i|\Lambda$. 

The two-point function is obtained by evaluating the geodesic length, 
\begin{equation}
\langle{\cal O}(x_f) {\cal O}(x_i)\rangle=e^{-m\int_{-T}^{T}d\tau}
=e^{-{4\over 5-p}JT}={1\over \left(\Lambda|x_f-x_i|\right)^{4J\over 5-p}}.
\label{eq:OOgeod}
\end{equation}
The exponent indeed agrees with the leading 
($J$-dependent) exponent in \eqref{eq:OOGKPW}
obtained by the GKPW prescription.

\paragraph*{Worldsheet analysis.---}
Null geodesic in the 10D spacetime, given by \eqref{eq:zt}
together with the $S^{8-p}$ part, is a classical solution of 
the closed-string worldsheet theory, 
in which the string is in a point-like 
configuration.
Fluctuations around the null geodesic can be studied by 
expanding the string action to the quadratic order around
the classical solution\footnote{%
In order to have Euclidean AdS and to keep the correct
number of physical degrees of freedom, we take 
one of the $S^{8-p}$ directions timelike. This prescription
has been introduced in \cite{DSY}, and is called double Wick rotation.}.
After eliminating unphysical modes by gauge fixing and 
the use of the constraints, we get eight bosonic and eight 
fermionic fields which are massive on the 
worldsheet~\cite{ASY, AS}\footnote{%
This analysis is very similar to the one performed by BMN~\cite{BMN}.
They study string (particle) moving around the center of Lorentzian AdS, 
and identity the AdS energy (in the global coordinates)
with the scaling dimension in the gauge theory. For $p=3$, our analysis 
based on the proposal of \cite{DSY} gives essentially the same results 
as BMN (see \cite{DSY, DY, DY2} for more discussion). But for $p\neq 3$
without conformal symmetry, the above procedure would not be
applicable, thus we should use our approach based on a string (particle)
that reaches the boundary.}.
The origin of their mass is the curvature of the
background spacetime.

The bosonic part of the quadratic action is~\cite{ASY,AS}\footnote{%
We have fixed the worldsheet metric as 
$\sqrt{h}h^{\tau\tau}=(\sqrt{h}h^{\sigma\sigma})^{-1}
=\tilde{r}^{(3-p)/2}(\tau)$~\cite{ASY}. 
With this choice, $m_x$, $m_y$ are constant, but $x'^2$, $y_i'^2$
get $\tau$ dependent coefficients. If we take the conformal
gauge~\cite{ASY}, $m_x$, $m_y$ would be $\tau$ 
dependent. Final result does not depend on the gauge choice.}
\begin{align}
 S^{(2)}&={1\over 4\pi\ell_s^2}\int d\tau\int_{0}^{2\pi\tilde\alpha} d\sigma
\Big\{\dot{x}_a^2+\tilde{r}^{p-3}(\tau)x_a'{}^2+m^2_x x_a^2\nonumber\\
&\qquad\qquad\qquad
+\dot{y}_i^2+\tilde{r}^{p-3}(\tau)y_i'{}^2+m^2_y y_i^2\Big\}, 
\end{align}
where the radius of the $\sigma$ direction is proportional to 
the angular momentum, $\tilde\alpha\propto J$~\cite{ASY,AS,HNSY2}.
The terms involving the spatial derivative 
$\partial_{\sigma}$ has a factor
$
\tilde{r}(\tau)\equiv 2\cosh\tau/\{ (5-p)\tilde\ell\} 
$
which depends on the position on the geodesic,
so the frequencies of string excited states are 
time dependent. 
Mass for the bosonic fields depends on whether the field come 
from the fluctuation along AdS$_{p+2}$ $(m_x)$ 
or $S^{8-p}$ $(m_y)$~\cite{ASY}, 
\begin{align}
m_x &=1 \qquad (p+1\mbox{ fields}),\nonumber \\
m_y &={2\over 5-p} \qquad (7-p\mbox{ fields}).
\label{eq:massb}
\end{align}
The quadratic action for fermions has been obtained similarly, 
starting from the Green-Schwarz action~\cite{AS}. 
Mass for the fermionic fields is\footnote{%
The gauge-fixed action has worldsheet supersymmetry only for $p=3$,
thus the mass of bosons and fermions are different for $p\neq 3$.}
\begin{align}
m_f &= {7-p\over 2(5-p)}\qquad (8\mbox{ fields}).
\label{eq:massf}
\end{align}

Ground state of the closed string corresponds to the 
operator \eqref{eq:OZJ}~\cite{BMN}.
The classical amplitude \eqref{eq:OOgeod} is corrected by contributions
from the fluctuations\footnote{%
The classical action for a massless particle 
vanishes, but \eqref{eq:OOgeod} is obtained by evaluating the
Routh function (where only an angular direction is Legendre
transformed from the Lagrangian to make $J$ an independent 
variable)~\cite{ASY}.}.
The frequencies of the lowest ($\sigma$-independent) modes are 
equal to the mass 
\eqref{eq:massb} and \eqref{eq:massf}. Their contribution to
the zero-point energy is
\begin{equation}
E_0={1\over 2}\left((p+1)m_x+(7-p)m_y-8m_f\right)=-{(3-p)^2\over 2(5-p)}.
\label{eq:E0}
\end{equation}
By including this correction\footnote{%
Contributions from higher modes (with wave
number $n\ge 1$ along $\sigma$) will cancel between 
fermions and bosons for $\ell_s\to 0$, since the
difference in \eqref{eq:massb} and
\eqref{eq:massf} is unimportant in this limit.}
to the classical result \eqref{eq:OOgeod},
we recover the GKPW result~\cite{AS}, 
\begin{equation}
 \langle  {\cal O}(x_f) {\cal O}(x_i)\rangle
=e^{-2 (m+E_0) T}={1\over (\Lambda |x_f-x_i|)^{{4J\over 5-p}+c_p}}.
\label{strong}
\end{equation}
with the value of $c_p$ given in \eqref{eq:c}.

\paragraph*{Free field result from the bulk.---}
Up to now, we have reviewed the framework for gauge/gravity correspondence
for D$p$-brane and the analyses expected to be valid at strong 't Hooft 
coupling. Assuming the validity of this framework, let us now consider 
what happens at zero 't Hooft coupling.

We make one additional assumption: we assume the operator 
\eqref{eq:OZJ} corresponds to a superstring whose worldsheet spatial 
direction is discretized into $J$ bits\footnote{%
Strings with discretized worldsheet appear in attempts 
to reformulate large $N$ gauge theory by 
Thorn~\cite{Thorn1, Thorn2, Thorn3}. It also appears in a proposal 
by Nielsen and Ninomiya~\cite{Nielsen1, Nielsen2} for
reformulating string theory by treating left and right movers
as independent. Direct relation of such work with ours is not clear 
at the moment.}.
This is natural if we recall the proposal by BMN~\cite{BMN}
(see also subsequent work~\cite{Verlinde1, Verlinde2, Zhou,
Bellucci, Danielsson, Alday})
that string excitations on the ground state \eqref{eq:OZJ} are 
represented in gauge theory by inserting ``impurities'' (fields $X_i$ 
with $i\neq, 8,9$, or derivatives $\partial_a$ with $a=0,1,\ldots, p$) 
into the sequence of $Z$'s. 

Weak 't Hooft coupling corresponds to strongly curved backgrounds.
In this limit, the string tension is much smaller compared
to the scale of the curvature of the background.  
If string tension is strictly zero (corresponding to zero
't Hooft coupling), binding force between bits does not exist,
and we can think of the string as a
collection of $J$ independent bits (particles). In this case, the
zero-point energy would be the sum of contributions from $J$ bits,
i.e. $J$ times $E_0$ that we obtained in \eqref{eq:E0}.
By including this correction, the correlator now becomes
\begin{equation}
 \langle  {\cal O}(x_f) {\cal O}(x_i)\rangle
=e^{-2 (m+JE_0) T}={1\over (\Lambda |x_f-x_i|)^{(p-1)J}}.
\label{strong}
\end{equation}
This is the free-field result:
mass dimension of a scalar field in $(p+1)$ dimension is $(p-1)/2$,
and the operator ${\cal O}$ consists of $J$ scalar fields.

\paragraph*{Discussion.---}
We have shown that the free-field result of the $(p+1)$-dimensional
maximally supersymmetric Yang-Mills theory can be reproduced 
from the bulk string theory for the two-point function of a 
particular operator ${\rm Tr}Z^J$. Our result is based on 
a natural assumption that the string corresponding to 
this operator is made of $J$ bits. We may regard this
as important circumstantial evidence 
that gauge/gravity correspondence works for general $p$ and
also at weak 't Hooft coupling. 

It remains to be seen whether our approach can be promoted to
a calculational framework applicable for finite 't Hooft 
coupling. (The issues mentioned below are not special to 
$p\neq 3$; they should be equally important for $p=3$.)
Having reproduced a zero 't Hooft coupling result form the bulk,
it is hoped that perturbative expansions of gauge theory in 
terms of 't Hooft coupling
can be reproduced as well\footnote{%
See \cite{Wadia, Alday, DY, DY2} for earlier work in that direction.}.
Discretized string action is not unique, thus it would  be
important to identify the interactions between bits
appropriately for this purpose.  

Let us make some comments to clarify the meaning of our proposal. 
First, in our proposal, we are assuming $N\to \infty$. 
In this limit in which we can ignore string loops, 
we should be able to treat 
the background to be fixed and non-fluctuating, even 
though it is 
strongly curved\footnote{%
In general, 
there could be $\alpha'$-corrections to the background, but we
expect the near-horizon D$p$-brane background is protected
against it due to supersymmetry, as is the case for AdS$_5\times S^5$.}. 
(In that sense, we are not really in the quantum gravity regime.)
Second, in the present work, we considered the large $J$ limit, 
which allowed us to perform a semi-classical study of string theory,
with its gauge symmetry completely fixed. It is an important
question how to formulate the correspondence at weak 't Hooft 
coupling for finite $J$, without semi-classical approximation
and gauge fixing, whether or not it is calculationally feasible.

A crucial question is whether one can construct 
a consistent string theory with discrete worldsheet. 
By discretization we may lose characteristic features of
string theory\footnote{%
See \cite{YoneyaSUP} for an illuminating account of string theory, 
with strong emphasis on general features and
physical interpretation.} such as the following ones.
First, open-closed string duality is due to the presence of 
infinite tower of modes on the world sheet. By discretization,
the closed-string level number is cut off at $n\lesssim J$.
This is not necessarily a problem, since the SYM theory 
consists of only the lowest modes of open string. It would be
very important to examine whether $n\lesssim J$ is just the
right number of degrees of freedom necessary for the duality 
in this context to hold. Second, the elimination of the 
UV divergence in string theory is due to modular invariance,
which is a remnant of conformal symmetry. It is not at all clear
what its counterpart is in a discretized theory. For our 
proposal at $N\to \infty$, the problem of string loop divergence 
is not relevant, but if we want to extend it to finite $N$,
this problem would be very important.

To gain insight for the above issues, it would be helpful to
extend our analysis to more general cases. One direction would
be to a class of operators with spins along the AdS$_{p+2}$ 
direction. These correspond to macroscopic strings rotating
in AdS~\cite{GKP2}\footnote{%
To the authors knowledge, this has not been studied for $p\neq 3$ 
(even for strong 't 
Hooft coupling); it would be necessary to reformulate the 
correspondence in Euclidean signature and look for strings 
that reach the boundary as is done in this work.}.
Another direction of extension would be to finite temperature. 
Finding the correct type of discretized string theory in this case 
may shed light on the Hagedorn behavior~\cite{AtickWitten}.

\paragraph*{Acknowledgements.---}
I would like to thank Tamiaki Yoneya for very helpful
discussions and comments. I also thank Tomotaka Kitamura and
Shoichiro Miyashita, Lenny Susskind for comments.

\end{document}